# Predicting user preferences

An evaluation of popular relevance metrics

*Pavel Sirotkin*

Heinrich Heine University Düsseldorf
Information Science Department
Universitätstr. 1
40225 Düsseldorf

sirotkin@phil-fak.uni-duesseldorf.de

**Abstract**

The many metrics employed for the evaluation of search engine results have not themselves been conclusively evaluated. We propose a new measure for a metric's ability to identify user preference of result lists. Using this measure, we evaluate the metrics Discounted Cumulated Gain, Mean Average Precision and classical precision, finding that the former performs best. We also show that considering more results for a given query can impair rather than improve a metric's ability to predict user preferences.

# 1. Introduction

One issue in the evaluation of information retrieval systems in general and search engines in particular is the missing benchmark for system evaluation metrics. Given the by now abundant number of measurement types, it is unclear how to judge them. With time, some measures like classical precision fall out of the researchers' favour, while new ones gain acceptance. However, the process is slow and by no means conclusive. As Büttcher et al. (2010, p. 410) note, "given their importance in IR evaluation, one might assume that the relationship between user satisfaction and, say, average precision has been thoroughly studied and is well understood. Unfortunately, this is not the case. User studies trying to find correlations between user satisfaction and various effectiveness measures are a relatively recent phenomenon."

Thus, the meaning of metrics is unclear. It might be conceivable that popular metrics measure user satisfaction; or user preference; or task completion; or success in reaching goals; or perhaps just the correlation with the metric itself and nothing else. We attempt to provide first answers to a particular question that has received relatively little attention until now: How well can popular metrics pick out user preference between result lists? We will describe some popular metrics and discuss previously conducted evaluations in Section 2. Section 3 describes our own methodology and introduces a new measure employed to produce results described in Section 4. Section 5 provides a discussion of those results, with conclusions drawn in Section 6.

# 2. Related Work

## 2.1. Evaluation metrics

A list by Della Mea et al. (2006), which does not claim any completeness, contains 45 evaluation metrics introduced from 1965 to 2005. Many of them have hardly been used except by their creators, and most are not relevant for our purposes. The earliest and surely most influential metric is precision. It is defined simply as the proportion of relevant documents returned by a system. Its shortcomings for the purposes of web evaluation have been often stated; particularly, as thousands of pages can be relevant to a query, it might be unwise to assign an equal weight to all returned results.

These concerns were partly addressed by Average Precision (AP). As its name suggests, it averages precisions at individual ranks. In words, AP considers the precision at every relevant result in the list, and divides it by the result's rank; then, the precision is averaged by dividing the sum of discounted precisions by the total number of relevant results. In most cases, the AP of many queries is considered, and a Mean Average Precision (MAP) is calculated (Formula 1). MAP is one of the most-used metrics and is employed in single studies as well as in large efforts such as TREC.

$$MAP = \frac{1}{|Q|} \left( \sum_{Q_i} \frac{1}{|R_i|} \left( \sum_{j=1}^{n} rel(D_j) \frac{\sum_{k=1}^{j} rel(D_k)}{r_j} \right) \right)$$

**Formula 1. MAP formula with queries *Q*, relevant documents *R*, documents *D* at rank *r* and *n* returned results. *rel* is a relevance function assigning *1* to relevant results.**

$$DCG_i = \begin{cases} CG_i & if\ i < b \\ DCG_{i-1} + \dfrac{G_i}{log_b(i)} & if\ i \geq b \end{cases}$$

**Formula 2. DCG with logarithm base *b* (based on Järvelin and Kekäläinen 2002)**

Another metric which has enjoyed wide popularity since its introduction is Discounted Cumulated Gain or DCG for short (Järvelin and Kekäläinen 2002). The more basic measure upon which it is constructed is the Cumulated Gain, which is a simple sum of the relevance judgements of all results up to a certain rank. DCG enhances this rather simple method by introducing "[a] discounting function [...] that progressively reduces the document score as its rank increases but not too steeply (e.g., as division by rank) to allow for user persistence in examining further documents" (Järvelin and Kekäläinen 2002, p. 425). In practice, the authors suggest a logarithmic function, which can be adjusted (by selecting its base) to provide a more or less strong discount, depending on the expectations of users' persistence (Formula 2). DCG can be modified to allow for better inter-query comparison; to this end, a perfect ranking for known documents is constructed. The DCG of a result list is then divided by the ideal DCG, producing normalized DCG (nDCG) in the 0..1 range.

## 2.2. Metric Evaluations

When a new evaluation metric is introduced, it is usually explained what its advantage over existing metrics is. Mostly, this happens in theoretical terms; more often than not, an experimental metric evaluation is also given. There are many studies comparing one metric to another; however, this has the disadvantage of being a circular confirmation, indicating at best correlation between metrics.

Another method was used for evaluating different CG metrics (Järvelin and Kekäläinen 2000; Järvelin and Kekäläinen 2002). Those were used to evaluate different IR systems, where one was hypothesized to outperform the others. The CG measures indeed showed a significant difference between the systems, and were considered to have been validated. We do not regard this methodology as satisfactory. It seems that evaluating the hypothesis with a new metric while at the same time evaluating the metric against the hypothesis may produce a positive correlation without necessarily signifying a meaningful connection to any outside entity.

More promising approaches attempt to judge metrics with regard to an external standard. These studies often cast doubt on assumptions about explicit measures. Several studies report that MAP does not correlate in a significant way with user performance (Hersh et al. 2000; Turpin and Scholer 2006). Another study showed some correlation (Kelly et al. 2007); however, it was significant for less than half of all users. Also, the study has methodological issues; it included only four topics, and, while raters were to formulate own queries, the result lists were predefined. That means that raters actually rated the same result lists for different queries. A further study examined the correlation between average precision and user success (Al-Maskari et al. 2008). The results showed a strong correlation between average precision and user success metrics (such as the number of retrieved documents) as well as user satisfaction. The correlation values are significant; however, the correlation was with a fourfold increase in average precision, which is quite an extraordinary difference. Compared with this distinction, the increase in user success and especially user satisfaction was quite low. When the (absolute or relative) difference between the systems' average precision was reduced, the significance of correlations promptly dropped and all but disappeared when the increase in average precision was at 30%. One more MAP study looked at average precision at rank 3, which was found to have a strong correlation with explicit user satisfaction (Huffman and Hochster 2007).

In a further study, precision, CG, DCG and NDCG were compared to three explicit measures of user satisfaction with the search session called "accuracy", "coverage" and "ranking" (Al-Maskari et al. 2007). The results were mixed. From the overall 12 relations between metric and user satisfaction, only two showed a significant correlation, namely, precision and CG with the ranking of results. There have been further studies indicating the need for more holistic ratings. Ali, Chang et al. (2005) have shown that the correlation between result-based DCG scores and result list scores (on a tertiary scale) is 0.54 for image and 0.29 for news search. While the fields were more specific than general web search, the numbers clearly do not indicate a reliable link between the scores.

These studies do not produce conclusive results, though they seem to cast doubt on the connections between popular metrics (as they have been used for web search evaluation) and user satisfaction. Therefore, the need for novel methods of metric evaluation has been emphasized (Mandl 2010).

## 3. Methodology

We attempt to provide a comparison of three popular explicit evaluation metrics in their relationship to user satisfaction. That is to say, we attempt to test whether and how well (M)AP and (n)DCG[1] indicate users' explicitly stated preferences. While there is no absolute standard against which to measure evaluation metrics, we consider user preference between two result lists to be a useful start. From the point of view of a search engine developer, the most interesting question to be answered by a metric is whether a given algorithm is better than another. This other might be a previous version of the algorithm, a competing search engine, or just a baseline value. Additional questions might regard the confidence in the preference statement or the amount of difference between the algorithms. And the most direct way to gather user preference is to obtain explicit judgments. The directness is needed to ensure that the standard we are measuring metrics against is not itself biased by an intermittent layer of theory. While a direct comparison of two result sets is not usual (and might be considered "unnatural" for search behaviour), we think it nevertheless provides a more close reflection of actual user preference than other methods.

For the study, the help of 31 first-year Information Science students was enlisted. They were required to enter queries they were interested in, as well as a detailed statement of their information need. For every query, the top 50 results were fetched from a major web search engine. From these, two result lists were constructed; one contained the results in original order, while the ordering of the other was completely randomized. Then the users were confronted, also through a web interface, with different types of judgments. First, they were presented with a query, an information need statement, and two result lists displayed side by side, which were anonymized and presented in random order. They were asked to conduct a search session as they would do normally, and when they were done, to indicate

---

[1] As we calculate the metrics on a per-query basis, nDCG is analogous to DCG while being easier to compare as it falls into the usual 0..1 range. Also, MAP for a single query is obviously equal to AP. For convenience, we will speak of MAP in all contexts.

$$PIR = \frac{\sum_{q \in Q} \begin{cases} sgn(m_{q1} - m_{q2}) * p_q & if\ |m_{q1} - m_{q2}| \geq t \\ 0 & else \end{cases}}{|Q|}$$

**Formula 3. Preference Identification Ratio with metric values *m*, queries *Q*, preference judgments *p* and threshold *t*.**

which result list they found better, or if both were equally good (or bad)[2]. Later, they were presented with single results and requested to evaluate their relevance given the query and the information need. Ratings were graded on a 1..6 scale, which is familiar to German students since it is the standard grade scale in schools and universities. For evaluation purposes, the ratings were converted to a 1..0 scale with 0.2 intervals (1 → 1.0, 2 → 0.8, …, 6 → 0.0). Both the preference and the relevance judgments could be for the users' own queries or for others'. The raters performed all actions via a Web interface.

The main evaluation measure was the ratio of queries for which the difference between metric values for the two result lists would correctly predict explicit user preference. We call the measure Preference Identification Ratio (PIR). The definition is given in Formula 3, with Q being the set of queries where the output of one algorithm has been judged to be better than another, $m_{q1}$ and $m_{q2}$ being metric values for the two result lists under comparison, $p_q$ the preference judgment (with value 1 if $q_1$ is preferred and -1 if $q_2$ is preferred), and *t* a threshold value to allow treating result list quality as equal if their metric values are similar. On an intuitive level, the numerator is the number of queries where we can correctly predict the user preference from explicit result ratings minus the number of queries where the preference prediction is inversed. The denominator is simply the number of preference judgments where a preference actually exists. If two result lists are judged to be of similar quality, a metric's values do not influence PIR, as choosing any one does not lead to any advantages or disadvantages to the user[3]. Were a metric to identify all preferences correctly, PIR would be 1; and if every preference judgment is reversed, PIR is -1. However, since assuming no preferences at all would result in a PIR value of 0, we can consider this to be the baseline.

## 4. Evaluation

Our aim was to determine how well MAP and nDCG predict user preference in different conditions as measured by PIR. For comparison, precision was also evaluated. In a departure from the classical definitions, we retained graded relevance values for precision and MAP. We defined Precision@K as the sum of relevance ratings at ranks 1 to K divided by K, which is a slight adjustment of the original formula also falling into the 0..1 range. For MAP, only the relevance function changes. The different conditions were different cut-off values, corresponding to different amounts of evaluation effort. If, after some rank, a further increase of the cut-off value provided only marginal PIR gains, one might lower the cut-off value and direct the released resources towards an increased number of queries.

---

[2] Interestingly (and surprisingly for us), the randomized result list was judged to be better than the original one in ca. 26% of all cases. The reasons for and implications of this finding go beyond the scope of this paper and will be discussed elsewhere.

[3] It may be argued that if the current algorithm performs equally well, the adoption of a novel one is a waste of effort. Here, though, we focus on user experience.

Our first task was to find appropriate values for the threshold *t*. To this end, we calculated PIR for every metric, cut-off value and threshold (the latter in 0.01 intervals); then, we selected thresholds performing best for every metric/cut-off combination. A sample *t* evaluation is shown in Figure 1.

While the PIR differences between neighbouring *t* values were expectedly small and thus not statistically significant, we feel justified in this approach as our main aim was inter-metric comparison. Thus, even if the better PIR of a threshold is due to chance, the influence of randomness should average out between thresholds, cut-off values and measures, and while the absolute numbers might be too high, the relative performance judgments are still relevant.

Now using the best available *t* values for every metric/cut-off combination, we were able to compare PIR performances. The relevant values can be seen in Figure 2. The first result that surprised us was the performance of precision. Rather than lagging behind the novel metrics, it outperforms MAP at lower cut-off values (1 to 6). The highest PIR value it reaches (0.8) is the same as the highest MAP value, and only slightly below the highest nDCG value (0.84). Also, we did not expect to see the performance of all three metrics decreasing after some cut-off value. While precision peaks at cut-off 6, nDCG at 7-8 and MAP at 8-9, the results strongly suggest that even within the cut-off values used here (which are significantly lower than those used in many evaluations such as TREC), less might be more.

A comparison of metrics and cut-off values suggests that in different circumstances, different metrics might be appropriate. MAP performs quite poorly at small cut-offs, but emerges as the best metric at 10. Precision never outperforms nDCG, but (at least at the earlier ranks) comes close enough for the difference to be minimal. In absolute terms, the maximum PIR reached is 0.84 (nDCG@7-8).

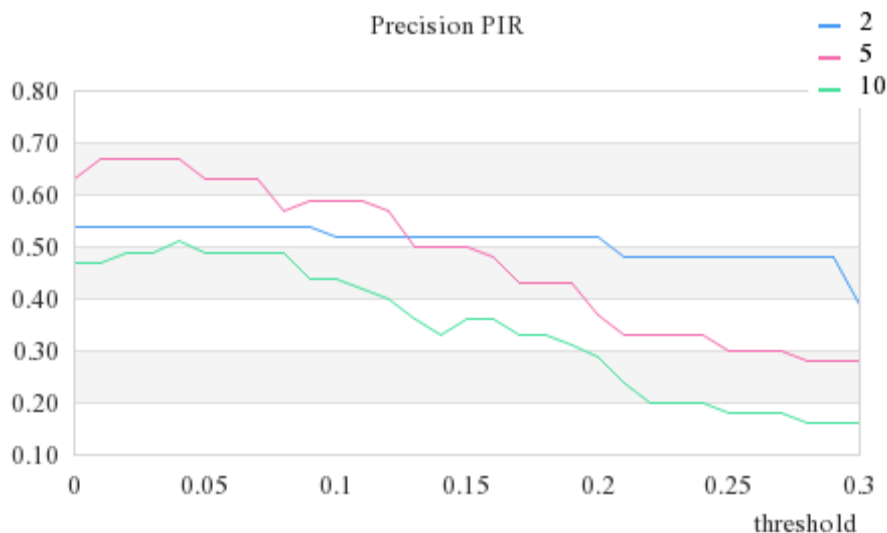

**Figure 1. Evaluation for different *t* values for precision with cut-offs 2, 5 and 10.**

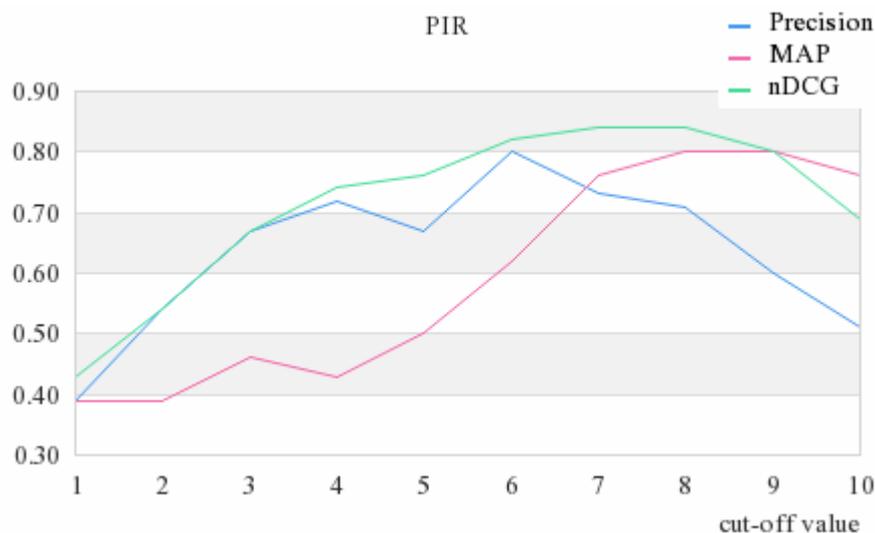

**Figure 2. PIR results**

## 5. Discussion

We would like to point out that search engine evaluation is just a small part of IR evaluation and, moreover, the type of performance we have attempted to capture is just one of many possible aspects of search engine quality. Lewandowski and Hochstötter (2007) propose a four-way quality framework including index quality, quality of the results, quality of search features and usability. The pure evaluation of organic, web page based result lists (as opposed to paid content or "universal search" features) is itself only a minimalistic subset of "quality of the results". However, the evaluated content is still an important and arguably even crucial part of a search engine's results. Also, our test subjects obviously did not constitute a representative sample of search engine users. While we look forward to studies with more diverse raters, the group is hardly less heterogeneous than those of most comparable studies.

Our results lead to some conclusions of practical importance. As an increasing cut-off value does not necessarily lead to a better approximation of user preferences, it might be a good idea to divert some resources from rating queries deeper to rating more queries. This has been found to provide higher significance (Sanderson and Zobel 2005); our results suggest that, rather than being a trade-off, exchanging depth for width can be doubly effective. It may even be sensible to reduce the cut-off to as low as 4, since it means cutting the effort in half while losing about 15% of information as measured by PIR. A possible explanation for the decrease of prediction quality is that users hardly look at documents beyond a certain rank (Hotchkiss et al. 2005), in which case any later difference in result quality is not reflected in actual user preferences. It would also explain why precision is the most and MAP the least affected, since the former has no and the latter a high discounting factor for later results.

Regarding individual metrics, nDCG was shown to perform best in most circumstances. In the best case, it correctly predicted 84% of user preferences. MAP might be employed if one explicitly desires to take into account later results, even if their relevance may not be important to the user. While precision performs considerably well, the present study has not found a situation where it would be the most useful metric.

The absolute PIR values we report may well be overestimations, as discussed in the Evaluation section. On the other hand, the preference judgments obtained were binary. We might assume that, given degrees of preference, we would find strong preferences easier to identify by considering document ratings. While metrics are often compared on their ability to distinguish between entities of relatively close quality, from the practical point of view, it is crucial for a metric to reliably pick out large differences, since those are the instances where the most improvements can be made. However, these conjectures await further research to confirm or disprove them.

Finally, our evaluation might have a value beyond its immediate results. We think that choosing an explicit, praxis-based standard for evaluating evaluation can help distinguish between the multitudes of available metrics. Particularly, a measure like PIR can be more practical than correlation measures often employed in such studies. Rather than indicating whether a given metric reflects a preference tendency, it can tell for what ratio of queries we would provide better results by using each metric to simulate preference judgments.

## 6. Conclusions and future work

A measure of a metric's ability to predict user satisfaction across queries was introduced. We used this measure, the Preference Identification Ratio (PIR), to provide estimates for the some common relevance metrics. (n)DCG was found to perform best, indicating the preferred result lists for up to 84% of queries. MAP provided good judgments at higher cut-off values, while precision did well without ever being the most informative metric. We also showed that search engine evaluations might be performed in a more significant and efficient way by considering more queries in less depth. The most significant cut-off values lie between 6 and 10 for different metrics, while the most efficient might come as low as cut-off 4.

Further work should look at PIR for degrees of preference and explore whether the metrics' performance stays at similar levels, and also to evaluate further metrics. We also intend to examine in more detail the cases where single metrics failed to pick a preferred result list to provide a qualitative analysis of their weaknesses. Finally, the connection of our results with log data might provide insights into relations between user behaviour and relevance or preference judgments.